\documentclass[sn-nature]{sn-jnl}

\usepackage{graphicx}%
\usepackage{multirow}%
\usepackage{amsmath,amssymb,amsfonts}%
\usepackage{amsthm}%
\usepackage{mathrsfs}%
\usepackage[title]{appendix}%
\usepackage{xcolor}%
\usepackage{textcomp}%
\usepackage{manyfoot}%
\usepackage{booktabs}%
\usepackage{algorithm}%
\usepackage{algorithmicx}%
\usepackage{algpseudocode}%
\usepackage{listings}%

\usepackage{amsmath}
\usepackage{hyperref}
\usepackage{caption}

\usepackage{todonotes}
\usepackage{tikz}
\usepackage{lipsum}
\usepackage{graphicx}
\usepackage{dcolumn}
\usepackage{bm}
\usepackage{verbatim}
\usepackage{tikz}
\usetikzlibrary{quantikz2}
\usepackage{multirow}
\usepackage{amsmath}
\usepackage{amssymb}
\newcommand{\sign}{\text{sign}}

\begin{document}
\title{Explainable quantum regression algorithm with encoded data structure}
\author*[1]{\sur{C.-C. Joseph Wang}}\email{josephwang13@gmail.com}
\author[2]{\sur{F. Perkkola}}
\author[2]{\sur{I. Salmenper\"{a}}}
\author[2]{\sur{A. Meijer-van de Griend}}
\author[2]{\sur{J. K. Nurminen}}
\author[1]{\sur{R. S. Bennink}}


\affil*[1]{\orgdiv{Quantum Computational Science Group, Quantum Information Science Section, Computational Sciences and Engineering Division}, \orgname{Oak Ridge National Laboratory}, \orgaddress{\city{Oak Ridge}, \postcode{37831}, \state{Tennessee}, \country{USA}}}

\affil[2]{\orgdiv{Department of Computer Science}, \orgname{University of Helsinki}, \state{Helsinki}, \country{Finland}}

\abstract{
Hybrid variational quantum algorithms (VQAs) are promising for solving practical problems such as combinatorial optimization, quantum chemistry simulation, quantum machine learning, and quantum error correction on noisy quantum computers.
However, with typical random ansatz or quantum alternating operator ansatz, derived variational quantum algorithms become a black box that cannot be trusted for model interpretation, not to mention deploying
as applications in informing critical decisions: the results of these variational parameters are just
rotational angles for the quantum gates and have nothing to do with interpretable values that a model can provide directly.
In this paper, we construct the first interpretable quantum regression algorithm, in which the quantum state exactly encodes the classical data table and the variational parameters correspond directly to the regression coefficients, which are real numbers by construction, providing a high degree of model interpretability and minimal cost to optimize due to the right expressiveness.
We also take advantage of the encoded data structure to reduce the time complexity of computing the regression map.
To shorten the circuit depth for nonlinear regression, our algorithm can be extended by building nonlinear features by classical preprocessing as the independent encoded column vectors. For measurable results by the algorithm, we show that model trainability
is achievable by success probability with the sample complexity that scales as $M\delta\epsilon(L, M)$, in which $L, M$ are
the respective number of rows and columns of the encoded structural classical data when the read-out error $\delta\epsilon(L, M)$ is the dominant error. For one-hot encoding, we show that the error $\delta\epsilon(L, M)$ is equal to $L(M+1)M\times\delta_{M}=N_{Q}M\times\delta_{M}$, in which $\delta_{M}$ is the read-out error for a physical qubit used for encoding, and $N_{Q} = L(M+1)$ is the number of allocated physical qubits in total. For compressed encoding, the read-out error $\delta\epsilon(L, M)$ equals $(N_{L}+N_{M})\delta_{M}$, exponentially suppressed (due to $N_{L}, N_{M}\ll N_{Q}$) than the one-hot encoding. Even though the realization of compressed encoding in superconducting qubits has been achieved by the less noisy compressed encoding recently by the authors, we envision potential quantum utilities with multi-qubit gates implemented in neutral cold atoms and ions.}
\maketitle

\section{Introduction}\label{sec1}
The interpretability and explainability of predictive models are essential for the wider adoption of machine learning and artificial intelligence applications, especially in domains where faulty model interpretation can have serious consequences. For example, in healthcare and financial applications, strict regulations require models with clear interpretation to validate model predictions, for the model to be approved/trusted. Model interpretability is an equally valid criterion for quantum machine learning, but so far has received little or no attention. Although a quantum regression algorithm was proposed a decade ago~\cite{quantum-algorithm-for-data-fitting, prediction-by-linear-regression-on-a-quantum-computer,fast-quantum-algorithms-for-least-squares-regression} for its fundamental importance and other approaches based on matrix inversion and quantum kernel methods have been proposed recently~\cite{Somma, Paine}, these works assumed noise-free quantum hardware and did not address model interpretation issues. In recent work with a hybrid variational ansatz to mitigate hardware noise~\cite{LANL}, as we did, they did not consider interpretation values for potential quantum applications.
We approach quantum regression from a variational perspective with a known encoded data structure and develop an algorithm that provides interpretive value and prediction power as required, useful in the noisy intermediate-scale quantum (NISQ) era~\cite{NISQ}.

Regression models are predictive models that learn the map between a target continuous variable and predictors (attributes/input variables/features) in training.
The predictor variables can generally be transformed into continuous variables with the appropriate interpretation based on the transformation performed. Regression models are important machine learning models to study due to their wider adoption in industrial applications at scale, as opposed to more complex models such as neural networks, which typically focus on predicted results and less on the descriptive correlation between the prediction outcomes and the predictors. Additional features, such as the flexibility to model nonlinear dependencies based on domain expertise and the ability to perform relevant variable selection with regularization techniques, further enhance the utility of regression modeling in statistical machine learning.

Variational quantum algorithms are undeniably the most feasible digital quantum algorithms to date~\cite{Variational quantum algorithms, A quantum approximate optimization algorithm, An adaptive variational algorithm for exact molecular simulations on a quantum computer, Hybrid Quantum-Classical Algorithms and Quantum Error Mitigation, Variational Circuit Compiler for Quantum Error Correction}. They offer a practical solution to bypass quantum hardware noises and intricate controls while maintaining their universality for quantum computation~\cite{Universal variational quantum computation}.
However, current hybrid variational algorithms (based on quantum alternating operator ansatz) are generic and approximate toward quantum applications, and therefore the direct connection of the variational gate parameters to interpretative parameters of values is often lost; we embark on the quantum regression problem differently in perspective by producing the exact regression map with the optimal gate parameters directly connected
to the regression coefficients crucial for the interpretability of the regression models.

The organization of the manuscript is as follows.
In Sect.~2, we motivate and introduce the abstraction of the full quantum algorithm, including amplitude encoding, and regression map generation with real variational parameters, which are explainable weights in classical regression problems and measurement. In Sect.~3, we present the quantum algorithm in Pauli spin language, which is pertinent for researchers working on quantum hardware. Therefore, the time complexity for the quantum algorithm can be analyzed naturally.
In Sect.~4, we simulate noisy measurement data gathered by quantum hardware with bootstrap sampling in conjunction with the ensemble of regression models to find the optimal weight parameters. We also show that regularization techniques are still a valid strategy for selecting important features in the context of the variational quantum regression algorithm, as in the classical regression algorithm.
In Sect.~5, we conclude with our findings.

\section{The explainable quantum regression algorithm}
\subsection{Problem statement}

Regression modeling is the task of determining the relationship between a set of independent quantities (or ``features'') $(X_1,\ldots, X_M)$ and a dependent quantity (or ``response'') $Y$ from experimental data. It is one of the most common and important tasks in science, with particular prevalence in data modeling and machine learning. Usually, the relationship is assumed to be linear in $X_m$, $Y = \sum_{m=1}^{M} W_m X_m$ where $(W_1,\ldots,W_m)$ are known as regression coefficients or importance weights.
However, by treating products of independent variables as additional independent variables, linear regression can also be used to model non-linear relationships. In the typical regression scenario, one has $L$ independent observations $(y_0,\ldots,y_{L-1})$ of $Y$ and the corresponding observations $(x_{0m},\ldots,x_{(L-1)m})$ of each variable $X_m$. The goal of (linear) regression is to determine the coefficients $(W_1,\ldots, W_m)$ that best fit the data.

We propose a new algorithm to solve the linear regression problem using variational quantum circuits, whose parameters encode the regression coefficients in a manner that allows for interpretable values. The best regression coefficients are found by classical optimization concerning a regularized cost function, which furthermore helps to find the subset of the most important features. A key aspect of our approach is that the structural data are encoded directly in the amplitudes of the quantum state, and the regression coefficients are encoded directly in the parameters of the quantum circuit, which leads to optimal interpretability. We note that protocols for implementing quantum amplitude encoding are still under active research ~\cite{Quantum state preparation protocol for encoding classical data into the amplitudes of a quantum information processing register's wave function, W state, TensorFlow Quantum: Impacts of Quantum State Preparation on Quantum Machine Learning Performance}. Along with our regression algorithm, we provide several state-preparation algorithms to facilitate the implementation of regression on near-term quantum computers.

\subsection{Quantum amplitude encoding}

The first step of our algorithm is to encode the observations $y_0,\ldots,y_{L-1}$, $x_{11},\ldots,x_{LM}$ in a quantum state.
For notational convenience, we define $x_{l0} \equiv y_l$ and define $\mathbf{X}$ as the matrix with elements $\mathbf{X}_{lm} = x_{lm}$ for $l=0,\ldots,L-1$ and $m=0,\ldots,M$.
We standardize the data by shifting and rescaling the data columns so that each column of $\mathbf{X}$ has zero mean and equal variance. This ensures that our algorithm is equally sensitive to all variables for the best training. The data is then globally normalized so that $\sum_{l,m} x_{lm}^2 = 1$.
Therefore, the data can, in principle, be mapped to the amplitudes of a quantum state:
\begin{equation}
|\psi_{D}\rangle = \sum_{l,m} x_{lm}|lm\rangle
\end{equation}
where $\{|lm\rangle\}$ are the computational basis states of a quantum system that has at least $L(M+1)$ orthogonal states.
For now, we do not discuss the details of possible encoding schemes or methods for preparing $|\psi_{D}\rangle$, as this would distract from the main ideas of the algorithm.
Details for physical implementation will be discussed in Sect. 3.

\subsection{Mapping of regression coefficients to quantum amplitudes}

Our goal is a variational circuit whose structure reflects that of the regression problem at hand and whose output is proportional to the regression error $E$ to be minimized with respect to $W_{m\in{\{1,\cdots,~M}\}}$
\begin{equation}
E = \sum_{l=0}^{L-1} ( y_l - \tilde{y}_l)^2,
\label{Eq:cost_function}
\end{equation}
in which
\begin{equation}
\tilde{y}_l = \sum_{m=1}^{M} x_{lm} W_m
\end{equation}
is the predicted value of $y_l$.

We show first how to multiply a given feature (column of $\mathbf{X}$) by a controllable coefficient.
It will be convenient for exposition to treat the row index $l$ and column index $m$ as separate quantum degrees of freedom, $|lm\rangle = |l\rangle \otimes |m\rangle \equiv |l\rangle |m\rangle$. Consider the operator
\begin{equation}
U^{m}(\phi)
= \boldsymbol{1} \otimes e^{-i\phi |m\rangle \langle m|}
\end{equation}
which acts as identity ($\boldsymbol{1}$) on the row (observation) register and imparts a phase to a selected element of the column (feature) register.  It maps $|l\rangle |m\rangle$ to $ e^{-i\phi} |l\rangle |m\rangle$ and leaves all other basis states unchanged. Thus, when applied to $|\psi_{D}\rangle$, it maps $x_{lm} \to e^{-i\phi} x_{lm}$ to all $l$. By extension, the sequence $\prod_{m=1}^{M} U^{m}(\phi_m)$ applies a controllable phase $\phi_m$ to each column $m$ of the data. In this case, the resulting state would be
\begin{equation}
|\psi_{D}\rangle = \sum_{l, m} x_{lm}e^{-i\phi_m}|l\rangle |m\rangle.
\end{equation}
Notice that the relation between $\phi_{m}$ and the coefficient of $|l\rangle |m\rangle$ is not exactly what we are looking for if we were to associate the phase $\phi_{m}$ with the real regression parameters. The quantum map would not be real (up to a global phase factor) and would not be linear in $\phi_{m}$ as expected for conventional linear regression.
Furthermore, the regression coefficients should range between $[-\infty, +\infty]$, while the unique range of $\phi_m$ is $[-\pi,\pi)$. Based on these observations, we cannot make a direct association of the phases $\phi_{m}$ with the regression weights $W_{m}$. However, if we engineer the circuit in a target code space to yield
\begin{equation}
|\psi_{l}\rangle  \propto \sum_{m} x_{lm} ( e^{-i\phi_m}+e^{+i\phi_m}) |l\rangle |m\rangle
\propto \sum_{m}x_{lm}\cos{\phi_m} |lm \rangle.
\end{equation}
We can identify $W_m \propto \cos{\phi_m} \in [-1.0, 1.0] $, with the proportionality chosen to bring the weights into the required range.

\subsection{Quantum regression algorithm}

To engineer this mapping of phases to regression weights, we use controlled phase gates of the form
\begin{equation}\label{eq: regression unitary}
U_{C}^{m}(\phi)
=|0\rangle \langle 0| \otimes \boldsymbol{1} \otimes e^{i\phi_m |m\rangle\langle m|} +
|1\rangle \langle 1| \otimes \boldsymbol{1} \otimes e^{-i\phi_m |m\rangle\langle m|}
\end{equation}
which act on an ancilla qubit for control, row register, and column register, respectively.
(Note that if the hardware does not natively support such a controlled gate with symmetric phases, it can be realized as an \emph{un}controlled rotation $e^{i \phi_m}$ followed by a controlled rotation $e^{-2i\phi_m}$ equivalently.)
This gate imparts the phase $ e^{i\phi_m}$ to $|0\rangle \otimes |l\rangle \otimes |m\rangle$, and leaves the states with column index $\neq m$ unchanged.
As we now show, the transformation $x_{lm} \to \cos\phi_m x_{lm}$ can be accomplished by such controlled phase gates with a suitably prepared and measured ancilla qubit. The steps of the algorithm and the corresponding evolution of the quantum state are as follows:
\begin{enumerate}
 \item Prepare the data state $|\psi_D\rangle$:
 \begin{equation}
     |\psi_{D}\rangle = \sum_{l,m} x_{lm}|lm\rangle
 \end{equation}
 (refer to physical implementation in Sect. 3).

 \item Prepare an ancilla qubit in the state $\ket{+} \equiv (\ket{0} + \ket{1})/{\sqrt 2}$:
 \begin{equation}
      \longrightarrow \ket{+} \otimes  |\psi_{D}\rangle.
 \end{equation}

 \item  Apply controlled phase gates $U_{C}^{m}$ for each column $m$:
 \begin{align}
     \longrightarrow & \quad \prod_{m} U_{C}^{m}(\phi_{m})
     \left( \frac{|0\rangle + |1\rangle}{\sqrt 2} \otimes  |\psi_{D}\rangle \right) \\
     & = \frac{1}{\sqrt{2}} \sum_{l,m} \left(  e^{i\phi_{m}} |0\rangle + e^{-i\phi_{m}} |1\rangle \right) \otimes x_{lm} |lm\rangle
 \end{align}

 \item Apply a Hadamard gate to the ancilla qubit:
 \begin{align}
 \longrightarrow & \quad \frac{1}{2} \sum_{l,m} \left( e^{i\phi_{m}} (|0\rangle + |1\rangle) + e^{-i\phi_{m}} (|0\rangle-|1\rangle) \right) \otimes x_{lm} |lm\rangle \\
 &=  \sum_{l,m} \left( \cos \phi_{m} |0\rangle + i \sin \phi_{m} |1\rangle \right) \otimes x_{lm} |lm\rangle
\end{align}

 \item Project the ancilla qubit onto the state $|0\rangle$:
 \begin{equation}
    \longrightarrow \quad |\Psi_{0}\rangle = \sum_{l,m} x_{lm} \cos\phi_{m} |lm\rangle
  \end{equation}

 \item Measurement by the hermitian operator:
   \begin{equation}\label{eq: measurement operator}
      \hat{M} = \sum_{l=0}^{L-1} \sum_{m,m'=0}^{M} |lm\rangle \langle lm'|.
   \end{equation}

\end{enumerate}

As shown in Appendix A, the expectation value $\langle \hat{M} \rangle \equiv \langle \Psi_{0}|\hat{M}|\Psi_{0}\rangle$ is
\begin{align}
 \langle \hat{M} \rangle  &= \sum_{l} \left( \sum_{m} x_{lm}{\cos\phi_m} \right)^{2}  \\
   & = (\cos\phi_0)^2 \sum_{l} \left( y_l - \sum_{m=1}^{M} x_{lm} W_m \right)^{2}
 \label{eqs:measurement1}
\end{align}
where we identify $W_m = -\cos\phi_m/\cos\phi_0$ as the regression coefficient for the feature $m$ ($M$
features in total)
and the response variable component $y_{l}$ is by definition the $x_{l0}$ component. With this identification, the sum in the equation above can be recognized as the regression error in Eq.~\eqref{Eq:cost_function}. This result bridges the gap between our quantum regression algorithm and the conventional regression algorithm and enables a clear interpretation of the variational parameters as discussed in our numerical studies.
In general, only the relative sign between the feature variables and the response variable matters, we can enforce condition $\phi_y \equiv \phi_{0} \in (\frac{\pi}{2}, \frac{3}{2}\pi)$ so that $\cos \phi_0$ is always negative and nonzero so that the regression coefficient $W_m$ is well defined. 
However, to simplify the quantum hardware implementation, we can restrict the response rotational angle $\phi_y$ to $\pi$ such that the projected probability measurement $\hat{M}$ leads exactly to the regression error (the mean squared error MSE), $\sum_{l} \left( y_l - \sum_{m=1}^{M} x_{lm} W_m \right)^{2}$ in classical regression problems with the simple regression coefficient $W_m = \cos\phi_m$. The MSE function is periodic and nonlinear in $\phi_m$.
To be explicit, we will still keep the response rotational angle $\phi_y \equiv \phi_0$ as a variational variable in later discussions.

\subsection{Model training and regularization}
Since the goal is to minimize the regression error, the simplest approach is to take the cost function $C(\mathbf{W})$ to be the mean squared error $MSE$,
\begin{equation}
 {\rm C}({\bf W}) = \sum_{l = 0}^{L-1}(y_l - \sum_{m = 1}^{M} x_{lm}W_{m})^{2} = \frac{\langle \hat{M} \rangle}{(\cos \phi_{0})^2}
\end{equation}
where the regression weights ${\bf W} = (W_1,\ldots,W_m)$ are implicit functions of the circuit parameters $\boldsymbol{\phi} = (\phi_0,\ldots,\phi_m)$.
The parameter vector $\boldsymbol{\bar{\phi}}$ that minimizes the cost function yields the optimal linear regression coefficients $\mathbf{\bar{W}}$ with $\bar{W}_m = -\cos\bar{\phi}_m / \cos\bar{\phi}_0 = \cos\bar{\phi}_m$.
To extract the row-local cost function $C({\bf W})$ by projected measurement $\langle M \rangle$ with gradient descent optimizer or parameter shift optimization, one can encounter barren plateaus~\cite{Barren plateau} in the cost function landscape. The gradient $\frac{\partial {\rm C}}{\partial \phi_{m}}=2\sin{\phi_{m}}\sum_{m'}\sum_{l}x_{lm}x_{lm'}\cos{\phi_{m'}}/\cos^{2}{\phi_{0}}$ for a feature $m$
, in which the quantum amplitude contraction $\sum_{l}x_{lm}x_{lm'}$
 is equal to the bounded Pearson's correlation coefficients $\sum_{l}x_{lm}^{c}x_{lm'}^{c}$ from the classical data with the superscript $c$, $|r_{m,m'}|=|\sum_{l}x_{lm}^{c}x_{lm'}^{c}|\in [-1,1]$, dived by the global normalization scaling factor $R^2$. 
The global normalization factor $1/R^2$ is proportional to $(M+1)^{-1}$ since the global normalization coefficient $R^{2} = M+1$ after typical classical data standardization with the mean subtraction and standard deviation division for all variables (including response variable, see Appendix B).
The data-agnostic barren plateau exists ($\frac{\partial {\rm C}}{\partial \phi_{m}} \rightarrow 0$) only with a large number of features (wide data table) $M \gg 1$
even when all the independent features considered are perfectly correlated with the response. (When the training data are noisy with negligible correlation with the response variable, a noise-induced barren plateau exists, consistent with the existence of the bad training data.)  
The hybrid algorithm does not encounter the typical barren plateau problem associated with a finite number of features considered in training, thanks to the row-local cost function we constructed. In the presence of noise, we can perform better model training using batched data from resampling data to mitigate error issues from measurement or cross-talk between physical qubits in the encoding, and potentially improve parameter-shift optimization in the presence of the noise-induced barren plateau by noisy training data.

A fundamental question to ask is how sensitive the cost function is to a well-trained model versus a poorly trained model
as the null (reference) model, in which the estimated response variable is the mean of the variable.
When the data used to train the model is noise-free and perfectly correlated with the response variable, we expect the cost function of a well-trained model to be zero by the cost function.  However, for the null (reference) model, the cost function is measurable with the corresponding success probability $Pr_{0}$. The minimal number of shots
with measurement error $\delta \epsilon$ is given by the inverse of $Pr_{0}(1-\delta \epsilon)$ (See Appendix~B).

In practice, the cost function in regression typically includes regularization terms to bias toward models that fit the data well with fewer features, which helps avoid overfitting~\cite{statistical machine learning} and mitigate the barren plateau from the important feature truncation from the regularization. The cost function is modified to the following:
\begin{equation}
{\rm C}({\bf W}) = \sum_{l = 0}^{L-1}(\sum_{m = 1}^{M} x_{lm} W_{m}-y_l)^{2} + \alpha\sum_{m = 1}^{M}|{W_m}| + \beta\sum_{m = 1}^{M}|{W_m}|^{2}.
\end{equation}
where $\alpha, \beta > 0$.  This cost function is given as a general elastic net regularization ($\alpha \neq 0,~\beta\neq 0$), which accommodates LASSO (least absolute shrinkage and selection operator, $L1:\alpha \neq 0$, $\beta = 0 $) or Ridge ($L2: \beta\neq 0$,~$\alpha = 0$) regularization as limiting cases. We note that the $L1, L2$ regularization terms can be evaluated on a classical computer and added to the cost function evaluated by the quantum computer.

With this general scheme, we can build our hybrid quantum-classical algorithm to find the best parameters $\phi_m,\alpha,\beta$ that minimize the overall cost function. We would tend to implement the popular gradient-based approaches with parameter shift to search the minima~\cite{Frans}. 
Instead, we explore a gradient-free algorithm, the Nelder-Mead (NM) optimization algorithm~\cite{Numerical Recipes}, for the cost function to search for global minima.
For current noisy hardware, model training with batched data is plausible. Moreover, with the gradual transition to fault-tolerant hardware, the strategy
with a parallelized NM algorithm with high-performance clusters is still valid with much larger batched data to process before ensemble
averaging. In this case, the gradient-descent-based approach is questionable due to serial
processing and potential barren plateau problems.
We found that convergence to the optimal value of the cost function to high accuracy can typically be found by passing the suboptimal result from the latest global NM search as the warm start parameters for the next global search iteratively until the desired accuracy is reached.
When model training involves a large amount of training data, training can be broken down into ensemble training with multiple bootstrap data samples in parallel.

\section{Implementation}

Previously, we have described how to implement linear regression in a variational quantum circuit with controlled phase gates. To have an end-to-end solution, we need to consider how to encode the data in the quantum state, that is, how to prepare the data state $|\psi_D \rangle$. To that end, we envision that the state can be prepared using programmable phase gates similar to those used to perform the regression. However, while in the regression step, the phases depended only on the feature and were the same for each observation, to encode the data, each distinct phase $\phi_{lm}$ will generally be needed for each data element $x_{lm}$.
We notice that due to the global normalization condition, each normalized element $x_{lm}$ is generally much smaller than 1. This indicates that we can encode these classical training data elements through small phase angles in which $\sin x_{lm} \approx x_{lm}$, that is, the phase angles are approximately the data elements themselves.

To minimize the potential hardware errors, we prefer low qubit counts while maintaining the simplicity of the algorithm. This is a particularly appealing solution for well-connected and programmable qubits such as Rydberg atom-based and ultra-cold ion quantum platforms~\cite{Martin, Martin2, Small Programmable Cold Ion}.
At this point, we consider the specifics of the data encoding.

\subsection{One-hot encoding}

\subsubsection{Data state preparation}

In one-hot encoding, each pair $(l,m)$ is mapped to a single index in $\{1,\ldots, L(M+1)\}$ and encoded by the value $1$ in the qubit with corresponding index.  This would require $L(M+1)$ qubits. One-hot encoding should be avoided for large data sets, as it requires more physical qubits.
For quantum machine learning on near-term quantum devices, this encoding is still useful
for the proof-of-concept of the algorithm we propose, since the circuits to implement it are relatively simple.

With one-hot encoding, the pair $(l,m)$ is mapped to a single index $j = m + l(M+1) \in \{0,\ldots,L(M+1)-1\}$.  The basis state $|lm\rangle$ is then encoded as $|1_j\rangle \equiv |0...010...0\rangle$, which has 1 for qubit $j$ and zero for every other qubit. This gives
\begin{equation}
    |\psi_{D}\rangle = \sum_{l,m} x_{lm} |lm\rangle
    = \sum_{j=0}^{L(M+1)-1} x_j |1_j\rangle
\end{equation}
The uniform superposition of one-hot-encoded states
is the well-known $W$ state that can be prepared by an efficient procedure~\cite{W state}.
The data state $|\psi_D\rangle$ can be prepared using essentially the same procedure but with modified rotation angles to produce the nonuniform amplitudes $x_j$.
The basic building block of this procedure is the gadget
\begin{figure}
\begin{center}
\begin{tikzcd}
& \ctrl{1} & \gate{X} & \qw \\
& \gate{R_{y}(\theta)} & \ctrl{-1} & \qw \\
\end{tikzcd}
\end{center}
\caption{Gadget for one-hot encoding \label{fig:one_hot_gadget}}
\end{figure}
in Fig. 1, consisting of a controlled-$\rm{Y}$ rotation followed by a controlled-NOT (CNOT) gate.
Such a gate can be realized without difficulty in many experimental platforms (for example, see Fig. 2 in the reference ~\cite{Small Programmable Cold Ion}).
This gadget maps $|10\rangle$ to $\cos\theta |10\rangle + \sin\theta|01\rangle$.
Starting with the state $|1_0\rangle = |10...0\rangle$ and applying this gadget with various angles to qubit pairs $(0,1), (1,2), (2,3), \ldots$ one can prepare an arbitrary superposition of basis states $|1_0\rangle, \ldots, |1_{L(M+1)-1}\rangle$. For the digital two-local gates, the run time complexity scales as $LM$ in the encoding.

\subsubsection{Quantum regression map}

In the one-hot encoding, the ancilla-controlled phase gate used to impart regression coefficients takes the form
\begin{align}
    U_{C}^{j}(\phi_j) &= \exp\left( -i \phi_j |1\rangle \langle 1| \otimes |1_j\rangle \langle 1_j| \right) \\
    & = \exp\left( -i \phi_j \frac{\rm{Z_{A}-I_{A}}}{2} \otimes \frac{\rm{Z_{j}-I_{j}}}{2} \right)
\end{align}
where $A$ denotes the ancilla qubit, $j$ indexes a data register qubit, $\rm{Z_A}$ ($\rm{Z_j}$) is the Pauli $\rm{Z}$ operator on qubit $A$ ($j$). It can be verified that $U_{C}^{j}(\phi_j)$ yields the desired effect on $x_j$ as
$U_{C}^{j}(|1\rangle \otimes |1_j\rangle) = \exp(-i\phi_j)(|1\rangle \otimes |1_j\rangle)$ and for every other state $U_{C}$ acts as the identity. In Table~\ref{table:full algorithm one-hot}, we summarize the full algorithm before measurement. The time complexity for the regression map is of $O(LM)$ for local gates but
can be further improved with non-local gates $O(M)$ similar to state preparation (See Appendix C).

\subsubsection{Measurement}

The measurement operator $\hat{M}$ is the summation of individual operators of the form $|lm\rangle \langle lm'|$.
This may be understood as a transition from $j=(l,m)$ to $j'=(l,m')$ which can be achieved by operators of the form $ \rm{S_{j'}^{+} S_{j}^{-}}$ where $\rm{S_j^{+}} = |1\rangle \langle 0| = (\rm{X_j - i Y_j})/2$ is the raising operator on qubit $j$ and $\rm{S_j^{-}} = |0\rangle \langle 1| = (\rm{X_j + i Y_j})/2$ is the lowering operator on qubit $j$. $\hat{M}$ can be written in terms of measurable quantities as
\begin{align}
\hat{M} &= \sum_{l}\sum_{m, m'}|lm\rangle\langle lm'| \\
 &= \rm{I} + \sum_{l}\sum_{m \ne m'} |lm\rangle\langle lm'| \\
& = \rm{I} + \sum_{l} \sum_{(j<k) = l(M+1)}^{l(M+1)+M}  \left( \rm{S_{j}^{+}S_{k}^{-} + S_{k}^{+}S_{j}^{-}} \right) \\
& = \rm{I} + \sum_{l} \sum_{(j<k)= l(M+1)}^{l(M+1)+M} \left( \rm{X_{j}X_{k}+Y_{j}Y_{k}} \right),
\end{align}
where $\rm{I}$ denotes the global identity (idle) operator. Thus $\hat{M}$ can be measured as a linear combination of $\rm{I}$, $\rm{X_j X_k}$, and $\rm{Y_j Y_k}$ measurements. Shadow tomography has been developed
to save the cost of performing the measurements effectively~\cite{Robert}. 

Take a two-by-two data table for encoding as an example, the basis states for $l=0$, $m=0,1$ are $|1000\rangle$ and $|0100\rangle$.
For $l=1$, $m=0,1$, the states are $|0010\rangle$ and $|0001\rangle$.
For $l=0$ the relevant state transition operators $\rm{{S_{0}}^{+}{S_{1}}^{-}}$ and $\rm{{S_{1}}^{+}{S_{0}}^{-}}$ are given by $|1000\rangle\langle 0100|$ and $|0100\rangle\langle 1000|$. For $l= 1$ the relevant state transition operators $\rm{{S_{2}}^{+}{S_{3}}^{-}}$ and $\rm{{S_{2}}^{+}{S_{3}}^{-}}$ are given by
$|0010\rangle\langle 0001|$ and $|0001\rangle\langle 0010|$.

The one-hot amplitude encoding introduced so far can be resource-intensive in qubits and error-prone, since the number of qubits scales with the number of classical data entries.
However, the overall quantum algorithm is relatively simple. For the near-term hardware, we expect one-hot encoding to be the easiest to implement for proof-of-principle demonstrations of our algorithm.  To mitigate hardware noise and reduce the size of quantum circuits needed, a batch training strategy is employed: First, we divide the training data into numerous batches of smaller bootstrap samples and train a regression model for each batch separately. Then the coefficients of the separate regression models are ensemble-averaged to produce the ensemble model, as will be demonstrated numerically in Sect.~{\ref{sec: numerics}}.

\begin{table}
\scalebox{0.90}{
\fbox{
\begin{tabular}{l}
{\bf Summary of Quantum Algorithm with One-hot Encoding} \\
\toprule
{\bf Input}: The data registry state $|\psi_{D}\rangle$ is initialized as $W$ state with one-hot encoding
\\
$|A\rangle \otimes |\psi_{D}\rangle  =
|+\rangle \otimes \sum\limits_{(l,m)} x_{lm}|lm\rangle $ \\
{\bf Output}:  The final quantum state before the projection measurement of the ancilla state $|0\rangle$:
$|A\rangle \otimes |\psi_{D}\rangle$ \\
$ = \sum_{l,m}x_{lm}\cos{\phi_{lm}}|0\rangle \otimes  |lm\rangle$
$ - ix_{lm}\sin{\phi_{lm}}|1\rangle\otimes|lm\rangle $\\
{\bf Procedure}:\\
(1) Initialization of the ancilla state $|+\rangle$ and $W$ states for the data registry $|\psi_{D}\rangle$. \\
(2) Application of multi-controlled phase gates with $\prod U_{C}^{m}$.\\
(3) Application of a Hadamard gate $H$ to the ancilla qubit $A$.\\
(4) Projective measurement conditioned on the ancilla state $|0\rangle$.
\end{tabular}
}
}
\caption{Summary of Quantum Algorithm with One-hot Encoding}
\label{table:full algorithm one-hot}
\end{table}

\subsection{Compact binary encoding}

For the one-hot amplitude encoding, the number of physical qubits allocated to support the information grows linearly with the number of data entries.
To extend the quantum algorithm in current noisy hardware, we need a much more compact
encoding scheme to minimize hardware noise due to a much larger qubit count for the same task.
Meanwhile, we want to again keep the structure of the classical data table and use the simple ancilla-controlled phase gates.

For this encoding scheme, the information from the rows and columns is stored in separate qubit register $|lm\rangle =|l\rangle \otimes |m\rangle$. $l$ and $m$ are encoded in binary using $N_L = \lceil \log_2 L \rceil$ qubits for $l$ and $N_M = \lceil \log_2 (M+1) \rceil$ qubits for $m$. That is, $\ket{l} = \ket{l_{N_L}} \otimes \cdots \ket{l_1}$ and $\ket{m} = \ket{m_{N_M}} \cdots \ket{m_1}$. Take, for example, a $4\times 4$ data table representing 4 observations each of 3 features and 1 response variable. In this case, the row indices $l=0,1,2,3$ would be represented by the four basis states $|00\rangle$,~$|01\rangle$,~$|10\rangle$, and $|11\rangle$, respectively; the same four basis states in the column register would encode $m=0,1,2,3$.

Thus, the number of qubits needed to store the entire data table is approximately $\log_2 L + \log_2 M = \log_2 LM$, which represents a substantial compression.
However, because of compressed encoding, the procedure to impart the regression coefficients into the quantum state has a much higher complexity in terms of one- and two-qubit operations.
Therefore, we will consider an alternative approach that exploits global entangling analog gates native to the latest cold-ion and Rydberg cold-atom systems.


\begin{table}
\scalebox{0.72}{
\fbox{
\begin{tabular}{l}
{\bf Summary of Quantum Algorithm with Compact Binary Encoding} \\
\toprule
{\bf Input}: The data registry state $|\psi_{D}\rangle$ is initialized as product state $\ket{\mathbf{X}}_{MEM}$ with binary encoding \\
for each data element indexed by a value binary string. \\
The ancilla qubit is initialized in $|+\rangle$ state and the QPU register is initialized in a uniform superposition state $|\Psi\rangle$ \\
with a computational basis encoding a unique key for each data element at a different location of the data table. \\
{\bf Output}:  The final quantum state before projection measurement of the ancilla state $|0\rangle$. \\
{\bf Procedure}:\\
(1) Initialization of the quantum state for the ancilla qubit, data register, and QPU register
$\ket{\Psi} = \ket{+} \otimes \left( \frac{1}{\sqrt{K}} \sum_{k} \ket{k}_{QPU} \right) \otimes\ket{\mathbf{X}}_{MEM}$\\
(2)  Data state preparation by multi-controlled phase gates $\prod U_{D}^{k}$ operating on data register
followed by \\ projective measurement on the ancilla state $|-\rangle$. \\
(3) Unitary rotation of the ancilla qubit from $|-\rangle$ state to $|+\rangle$ state.\\
(4) Quantum regression map generation by multi-controlled phase gates $\prod U_{C}^{m}$ operating on QPU register. \\
(5) Application of a Hadamard gate $H$ to the ancilla qubit $A$.\\
(6) Projective measurement on the ancilla state $|0\rangle$.
\end{tabular}
}
}
\caption{Summary of Quantum Algorithm with Compact Binary Encoding}
\label{table:full algorithm binary}
\end{table}

\begin{figure}[h]
    \centering
    \includegraphics[scale=0.35]{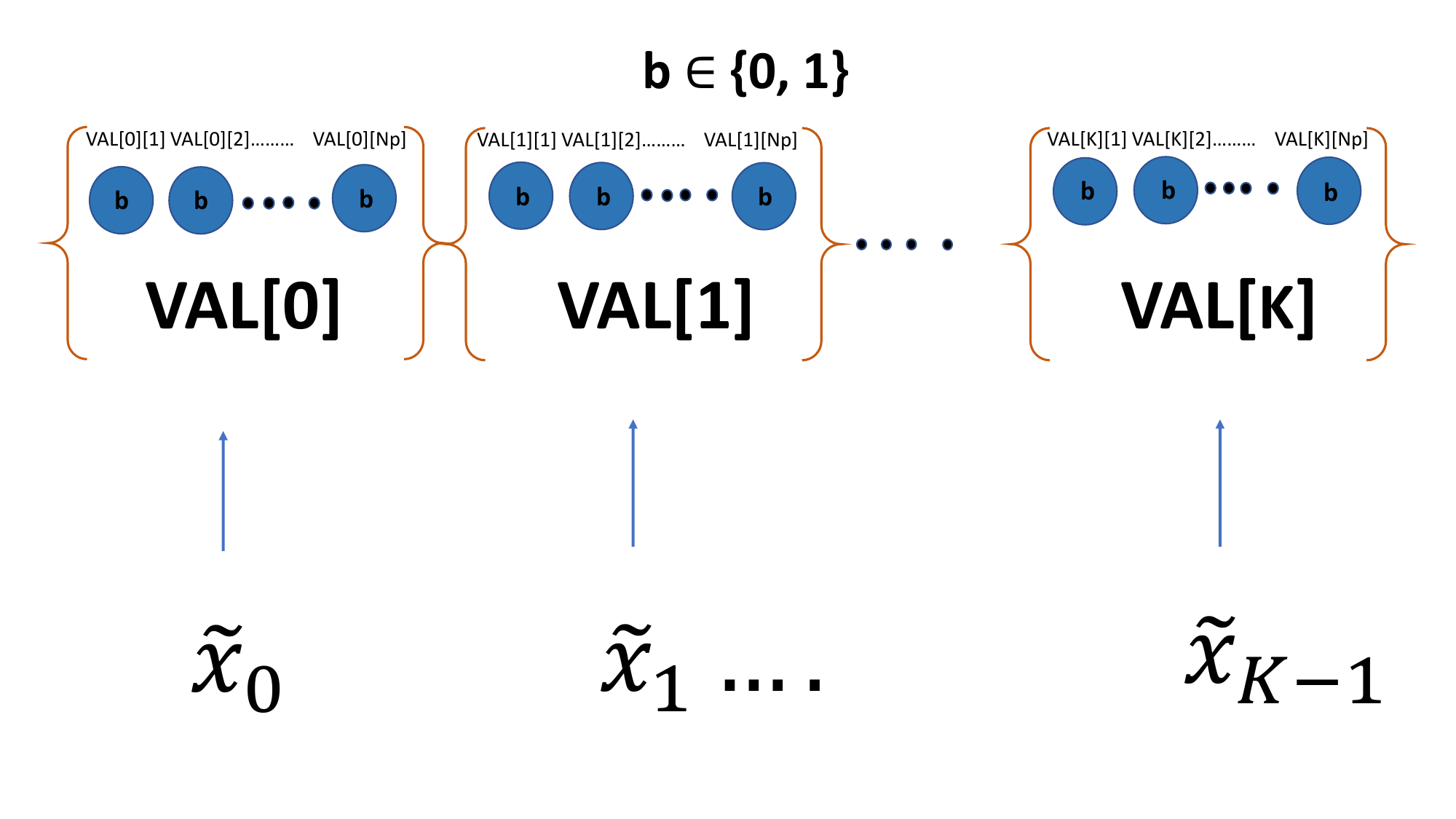}
    \caption{Quantum memory registry resource allocation: a physical qubit is represented by a filled circle
    with a binary code $b \in~\{0,~1\}$.}
    \label{fig:QRA}
\end{figure}

\subsubsection{State preparation}

To prepare a quantum state $|\psi_D\rangle$ containing the classical data $\bf X$ in a binary encoding scheme, we consider the scheme in \cite{Quantum state preparation protocol for encoding classical data into the amplitudes of a quantum information processing register's wave function} in which the real data is first digitized and programmed into a computational basis state supported by the quantum memory register. This needs only to be done once upfront. Subsequently, each time a copy of the quantum data state $|\psi_D\rangle$ is initialized, it is prepared by a known and efficient circuit that coherently applies phases stored in the quantum memory register to a quantum processing unit (QPU) register, without destroying the data in the memory register. The quantum information in the memory register can also be reinitialized when the quantum coherence time is surpassed without changing the quantum algorithm in the quantum processing unit.

We assign the index $k=0,\ldots, K-1$ as the entries of a data table, where $K = L(M+1)$.  Each data element $x_k$ is digitized as $\tilde{x}_k$ and stored in a separate qubit register (See Fig.~\ref{fig:QRA}):
\begin{equation}
 \ket{\tilde{x}_k}_{VAL[k]}.
\end{equation}
To digitize $x_k$, let $a$ be an upper bound on the magnitude of the data: $\max_k |x_k| < a$, in which the magnitude of $a$ depends on the data table that is standardized and globally normalized before encoding. Then $x_k \in [-a,a]$ is approximated using $N_P$ bits of precision as
\begin{equation}
 x_k \approx \tilde{x}_k = a \left( 2^{-1} (-1)^{x_{k,1}} + 2^{-2} (-1)^{x_{k,2}} + \cdots + 2^{-N_P} (-1)^{x_{k,N_P}} \right)
\end{equation}
where $x_{k,1},\ldots,x_{k,N_P} \in \{0,1\}$.  $\tilde{x}_k$ is then stored in the memory as
\begin{equation}
 \ket{\tilde{x}_k}_{VAL[k]} = \ket{x_{k,1}} \ket{x_{k,2}} \cdots \ket{x_{k,N_P}}.
\end{equation}
The full state of the memory register is
\begin{equation}
    \ket{\mathbf{X}}_{MEM} = \bigotimes_{k=0}^{K-1}  \ket{\tilde{x}_k}_{VAL[k]}.
\end{equation}
The total number of qubits for the memory register is $K N_P \approx LM \log_2 (\epsilon^{-1})$ where $\epsilon = 2^{-N_p}$ is the precision of each data element.
While the number of qubits is linear in the size of the data table, as mentioned previously, these qubits need only be kept in a classical digital state.

Once the data has been stored in the memory register, a fixed circuit uses the memory register coherently to ingest the discrete data to the amplitudes of the superposition state $|\psi_D\rangle$ on the QPU register. We introduce an ancilla qubit in the state $\ket{+}$ and an $N_K$ qubit QPU register in a uniform superposition of all the binary encoded keys, yielding the state
\begin{equation}
     \ket{\Psi} = \ket{+} \otimes \left( \frac{1}{\sqrt{K}} \sum_{k} \ket{k}_{QPU} \right) \otimes\ket{\mathbf{X}}_{MEM},
\end{equation}
in which the state $|k\rangle_{QPU}$ is the shorthand for the encoded key $|l\rangle |m\rangle$ for the data location in the table. Note that the QPU register size $N_K =\log_{2}K \approx \lceil \log_2 (L) \rceil + \lceil \log_2 (M+1) \rceil\approx \log_2 (LM)$ is much smaller than that of the (classical) memory register, which reduces the opportunities for hardware errors.
The essential step is a unitary that transfers the digitized classical data $\ket{\tilde{x}_k}$ to the phase of the ancilla qubit when the key in the QPU register is $k$ \cite{Quantum state preparation protocol for encoding classical data into the amplitudes of a quantum information processing register's wave function}:
\begin{equation}{\label{eq:state_preparation}}
U^{k}_{D} = \exp \left( -i \rm{Z_A} \otimes   \ket{k}\bra{k}_{QPU} \otimes \hat{\theta}_{k} \right)
\end{equation}
The operator $\hat{\theta}_{k}$ is given by
\begin{equation}\label{eq:theta_k}
    \hat{\theta}_k = \sum_{j=1}^{N_P} \Delta \theta_j \rm{Z_{k,j}}
\end{equation}
in which $\rm{Z^{k}_{j}}$ is the operator on the $j$-th qubit on the value register $|VAL\rangle[k]$. When applied to the memory register, $\hat{\theta}_k$ is evaluated to the $N_P$-bit approximation to $x_k$:
\begin{equation}
\hat{\theta}_{k} \ket{\mathbf{X}}_{MEM} = \tilde{x}_k \ket{\mathbf{X}}_{MEM}.
\end{equation}
Notice that the phase $\Delta \theta_j = a 2^{-j}$ is predetermined and can be realized by programming quantum gates with suitable gate times and interaction strengths, although fully programmable quantum hardware on a large scale is still an active research area in hardware implementation~\cite{Small Programmable Cold Ion, Monroe}.
The factor $\ket{k}\bra{k}_{QPU}$ in the exponent causes this phase to be produced only if the state of the QPU register matches the key $K$. Explicitly,
\begin{equation}
    U_D^k \left( \ket{b}_A \otimes \ket{k'}_{QPU} \otimes \ket{\mathbf{X}}_{MEM} \right) = e^{-i (-1)^{b} \tilde{x}_k \delta_{k,k'}} \left( \ket{b}_A \otimes \ket{k'}_{QPU} \otimes \ket{\mathbf{X}}_{MEM} \right)
\end{equation}
where $b \in \{0,1\}$. Then
\begin{align}
    \prod_k U_{D}^{k} \ket{\Psi} &=  \left(  \frac{1}{\sqrt{K}} \sum_{k} \frac{ e^{-i \tilde{x}_k } \ket{0}_A + e^{i \tilde{x}_k}  \ket{1}_A }{\sqrt{2}} \otimes \ket{k}_{QPU} \right) \otimes\ket{\mathbf{X}}_{MEM}.
\end{align}
The encoded data state is then realized by projecting the ancilla qubit onto $\ket{-}$:
\begin{align}
    \bra{-}_A \prod_{k} U_D^{k} \ket{\Psi}
    & \propto  \sum_{k} \sin \tilde{x}_k \ket{k}_{QPU} \otimes  \ket{\mathbf{X}}_{MEM} \\
    &\approx \ket{\psi_D}_{QPU} \otimes  \ket{\mathbf{X}}_{MEM}
\end{align}
since $\sin \tilde{x}_k \approx \tilde{x}_k \approx x_k$ for a standardized data table.

The unitary $U_D = \prod_k U_D^{k}$ is rather complex with terms involving many-qubit Pauli operators. Here we show that we can take advantage of nonlocal M{\o}lmer-S{\o}rensen (MS) gates, which are available in current cold-ion technology~\cite{Peter Zoller, Peter Zoller2} and an active research area in Rydberg-atom platforms~\cite{Martin, Martin2}. We first expand the key selection operator in terms of Pauli strings and the binary encoding of $k$  as $k = k_{N_K} \cdots k_2 k_1$:
\begin{align}\label{eq:kk expansion}
   \ket{k}\bra{k}_{QPU} &= \prod_{i=1}^{N_K}
   \frac{ \mathbf{1} + (-1)^{k_i}\rm{Z^{QPU}}_{i}}{2} \\
   & = 2^{-N_K} \sum_{P \in \{ \rm{I,Z} \}^{\otimes N_K}} (-1)^{\frak{p}(\rm{P})} \rm{P_{QPU}},
 \end{align}
Here $\frak{p}(\rm{P}) \in \{0,1\}$ is the parity of these bits of $k$ that correspond to factors of $\rm{Z}$ in $\rm{P}$.
As a result,  $U_{D}^{k}$ can be written as $ U_{D}^{k} = \prod_{j=1}^{N_P} U_{D}^{k,j}$ where
\begin{equation} \label{eq:U_kj}
     U_{D}^{k,j} = \prod_{\rm{P} \in \{ \rm{I,Z} \}^{\otimes N_K}}
    e^{-i 2^{-N_{K}} (-1)^{\frak{p}(\rm{P})} \Delta \theta_j \rm{Z_A} \otimes  \rm{P_{QPU}} \otimes \rm{Z_{k,j}}}.
\end{equation}
Notice that each factor in $U_{D}^{k,j}$ is a multi-qubit Pauli rotation, where the operator in the exponent is a product of Pauli $\rm{Z}$ operators operating on selected qubits.
It will soon be possible to implement such rotations efficiently in fully programmable cold ion or cold atom qubit architectures. As discussed in \cite{Peter Zoller2} and Appendix C, a many-qubit rotation can be realized by a short (length $O(1)$) sequence of nonlocal M{\o}lmer-S{\o}rensen (MS) gates in conjunction with one-qubit ancillary gates on selected qubits. A basic MS gate operation generates a global set of pairwise interactions, while ancilla qubits in conjunction with MS gates generate interactions for Pauli strings on as many qubits as needed. We point out that this is an example of rarely-discussed digital-analog quantum computation~\cite{Eugene}.
In any case, the time complexity to implement $U_{D}^{k}$ using such an approach is $O(N_P 2^{N_K}) \approx N_P LM$.
There are $LM$ keys, so the overall time complexity for state preparation $\prod_{k}U_{D}^{k}$ is $LM N_P 2^{N_K}\approx N_{P}(LM)^2$.
Notice that the state preparation is more demanding due to the quantum data being injected by the quantum memory registry in comparison with the one-hot-encoder introduced earlier, where the classical resources are used.
If we replace the quantum memory registry with the classical resource ($\theta_{k}$ instead of ${\hat \theta}_{k}$ in Eq.~(32)), the gate complexity $U_{D}^{k}$ can be further reduced to $O(LM)$ and the time complexity for the state preparation is of $O(L^2M^{2})$.

In comparison, the cost of implementing $U_{D}^{k,j}$ with digital local gates is greater. By the discussion of Hamiltonian simulation on page 210 of \cite{book}, the time complexity to implement a multi-qubit rotation using local digital gates is roughly proportional to the number of qubits involved.  Thus the time complexity to implement $U_{D}^{k,j}$ using local digital gates is on the order of $\sum_{n=1}^{N_K} n \binom{N_k}{n} = N_K 2^{N_K}$ with
the overall time complexity $\prod_{k,j}U_{D}^{k,j}$ estimated as $LM N_{P} N_K 2^{N_K}$. This is greater than the time complexity of the suggested global MS implementation by a factor of $N_K \approx \log_2 (LM)$.

\subsubsection{{\bf Quantum regression map}}

To impart the regression coefficients into the data state $\ket{\psi_D}$ the memory register is not needed; the coefficients are imparted by the unitary $U_{C}^{m}$, Eq.~(\ref{eq: regression unitary}), acting on the QPU and ancilla register:
\begin{equation}
    U_{C}^{m} = e^{i\phi_{m} \rm{Z_A} \otimes \mathbf{1} \otimes \ket{m}\bra{m}}.
    \label{eq:reg}
\end{equation}
$U_{C}^{m}$ is analogous to $U_{D}^{m}$ but with two main differences. First, while $U_{D}^{k}$ selects a specific data element $k=(l,m)$, $U_{C}^{m}$ selects only the column $m$ and performs identically on each row $l$ of the data table.  The second difference is that, while the phase imparted by $U_{D}^{k}$ is encoded digitally in the quantum VAL register, the phase appearing in $U_{C}^{m}$ is a simple scalar determined by the regression coefficient.

$U_{C}^{m}$ can be implemented using the same strategy as $U_{D}^{k}$.  Recall that the column index $m$ is represented in binary as $m = m_{N_M} \cdots m_2 m_1$ where $N_M = \lceil \log_2 (M+1) \rceil$. Then
\begin{equation}
    \ket{m}\bra{m} = \otimes_{j=1}^{N_M} \ket{m_j}\bra{m_j} = \prod_{j=1}^{N_M} \frac{\mathbf{1} + (-1)^{m_j} \rm{Z_j}}{2}
\end{equation}
where $Z_{j}$ is the operator on $j$-th qubit in the $\ket{m}$ register within the QPU register. Upon factoring, the product $U_C^{m}$ may be written as
\begin{equation}
     U_{C}^{m} = \prod_{\rm{P} \in \{\rm{I,Z}\}^{\otimes N_M}} e^{+i(2^{-N_{M}}\phi_{m} (-1)^{\frak{p}(\rm{P})} \rm{Z_{A}}\otimes {\bf 1} \otimes \rm{P_{QPU}}},
\end{equation}
where this time $\frak{p}(\rm{P})$ is the parity of those bits of $m$  that correspond to factors of $\rm{Z}$ in $\rm{P}$.  Again, these multi-qubit rotations can be implemented either as a multi-qubit controlled gate or using multi-qubit  M{\o}lmer-S{\o}rensen gates as discussed above.
The time complexity for the feature mapping in quantum regression
scales as $2^{N_{M}}\times (M+1) \approx  M^{2}$.
In Table~\ref{table:full algorithm binary}, we summarize the full algorithm before the measurement.
In Appendix C, we discuss potential hardware implementation and resources for gate operation for interested readers.

\subsubsection{Measurement}
In the binary encoding, the measurement operator $\hat{M}$, Eq.~(\ref{eq: measurement operator}) takes a particularly simple form:
\begin{equation}
    \hat{M} = \sum_{l=0}^{L-1} \ket{l} \bra{l}
    \sum_{m,m'=1}^{M} \ket{m} \bra{m'}.
\end{equation}
Using the binary expansion of $\ket{l}$ we have $\sum_l \ket{l} \bra{l} = I^{\otimes N_L}$. Similarly, $\sum_{m,m'} \ket{m} \bra{m'} = (2 \ket{+}\bra{+})^{\otimes N_M} = (I+X)^{\otimes N_M}$. Thus
\begin{align}
    \hat{M} &= 2^{N_M} \rm{I}^{\otimes N_L} \otimes (\ket{+} \bra{+})^{\otimes N_M} \\
    & = \rm{I}^{\otimes N_L} \otimes (\rm{I+X})^{\otimes N_M}
\end{align}
Take a 2-by-4 data table, for example. The $\ket{k}$ states for $l=0$ are $\ket{0}\ket{00}$, $\ket{0}\ket{01}$, $\ket{0}\ket{10}$, and $\ket{0}\ket{11}$; the states for $l=1$ are analogous.  In this case
\begin{align}
    \hat{M} &= \rm{I} \otimes (\rm{I+X}) \otimes (\rm{I+X})
\\
 &= \rm{I \otimes I \otimes I + I \otimes I \otimes X + I \otimes X \otimes I + I \otimes X \otimes X}.
\end{align}
$\ket{\Psi_0} = \psi_{000} \ket{0}\ket{00} + \cdots + \psi_{111} \ket{1} \ket{11}$ denotes the state of the $QPU$ register just prior to measurement.  This reproduces the projected probability measurement results for all rows as 
\begin{align}
    \langle \hat{M} \rangle &=
    |\psi_{000} + \psi_{001} + \psi_{010} + \psi_{011} |^2 +
    |\psi_{100} + \psi_{101} + \psi_{110} + \psi_{111} |^2.
\end{align}

\section{Numerical results}
\label{sec: numerics}
Conventionally, to draw reliable interpretations from a trained regression model, we need to characterize the statistics of the uncertainty for the corresponding regression parameter for the predictor variables to justify its relevance in explaining the data. Motivated by the bootstrap aggregation (bagging) and the success of the random forest algorithm, we can build a regression model with bootstrap samples and compute the average and the standard errors (SEs) of the predicted regression coefficients from the ensemble of corresponding regression models by drawing the same number of bootstrap data samples from the original master (data) population (See reference ~\cite{bootstrap1, bootstrap2, Statistical Learning} for bootstrap sampling concepts and numerical analysis). The approach is supported theoretically~\cite{BoLasso}. 
Since the qubits in quantum hardware would be noisy to handle a large data set, a plausible solution is to train the regression model from smaller bootstrap samples from the smaller subsets of the training data with the same circuit to quantify errors and gather the final bootstrap statistics of the regression parameters by averaging the measurement results from the quantum algorithm running by the quantum hardware. 
Because of the exact mapping of classical regression models into quantum ones in our proposal, the statistical properties for the classical regression model still apply to the hybrid quantum regression model, as
illustrated in the following numerical demonstration as an example. The generalization and extension to other quantum machine models need to be explored further.

Here we show the promise of quantum-encoded data that can be processed in well-connected quantum hardware and provide an alternative hybrid quantum solution for quantum machine learning applications. For the proposed variational quantum regression (VQR), we show a different and robust strategy to use a global optimization search algorithm to find the optimal regression coefficients to avoid measurement overheads based on gradient-based approaches.
The best estimation can be found by using the suboptimal solutions with lower accuracy for regression coefficients as a new ansatz initialization for the next round of the global Nelder-Mead (NM) optimization algorithm until the final converged solution to high accuracy is found. For numerical demonstration, we adopt NM optimization algorithm from SciPy ~(an open-source Python library for scientific and technical computing) to validate the batch learning strategy with the analytically known cost function in Eq.~(17) (for larger data applications with distributed bootstrap samples, the distributed NM optimizers can be used).
In the following numerical results, the tuning variables for the cost function are cosine functions for the phase angles
instead of the phase angles.
The search for our optimal solutions is more effective with the new variables because of the unconstrained search
for the NM optimizer.

\subsection{Ensemble model training}
Machine learning from an ensemble model can be useful for statistical modeling, so that it is scalable with large data.
We trained an ensemble model from $N_b$ sets of bootstrap samples of various sizes. The best model is determined by the estimated weight vector ${\bf \widetilde {W}}$ calculated from the estimated feature weights ${\bf {\widetilde W}}=({\widetilde W}_{1}, \widetilde{ W}_{2}, ...., \widetilde{W}_{M})$ from bootstrap samples, that is, ${\widetilde{W}_{i}} = N_{b}^{-1}\sum_{b = 0}^{N_b-1}{\widetilde{W}}_{i}^{b}$ in which $\widetilde{W}_{i}^{b}$ is the weight learned from the batch $b$ for the feature $i$ and the standard errors (SEs) for the weights from model training is denoted as $\delta{\widetilde W}_{i}$.
To validate the ensemble learning, we generate synthetic and standardized classical data sets with a deterministic linear map with small randomness between the $M$ features $X_{j} = (X_{j,1}, X_{j,2},..., X_{j, M})$ and the target variable $Y_{j} \in \mathcal{R}$.
Specifically, the ideal (noiseless) linear map is given by the expression
$Y_{j} =  X_{ji} \overline{W}_{i}$ where $X_{ji}$ is the $L$-by-$M$ data matrix and
the best weight vector $\overline{\bf W} = (\overline{W}_{1},\overline{W}_{2},...,\overline{W}_{M})$ of size $M$ after data standardization.
We let each feature follow the uniform random distribution between values $[-1,1]$ to cover the feature space. 
For the response column $Y_j$, it is generated by the linear map $Y_{j} = {X_{ji}}W_{i}$ with the random variables $\{W_{i = 1,2,.., M}\}$ with the ideal population mean ${\bf \overline W}=({\overline W}_{1}=1.0,{\overline W}_{2}=2.0,{\overline W}_{3}=3.0,..,{\overline W}_{M}=float(M))$ and the standard deviation $\delta$(the same for each feature) from its mean value ${\overline W}_{i}$.
We would expect the model training would be more uncertain for the first few features due to the smaller signal-to-noise ratio ${\widetilde W}_{i}/\delta \widetilde{W}_{i}$ where $i = 1,2,..., M$ from an equal number of bootstrap samples with different sample sizes.
Notice that the weights ${\widetilde{W}}_{i}$ and the SEs $\delta{\it W}_{i}$ from training are in tilde to differentiate from the mean weight ${\overline W}_{i}$ and the standard deviation $\delta$ from the data generation respectively.

\begin{table}[h]
\begin{center}
\scalebox{0.90}
{
\begin{tabular}{||c|c|c|c|c|c|c|c|c|c||}
\hline
Sample size & \scriptsize ${\widetilde W}_{1}$ & \scriptsize ${\widetilde W}_{2}$ & \scriptsize ${\widetilde W}_{3}$ & \scriptsize ${\widetilde W}_{4}$ & \scriptsize ${\widetilde W}_{5}$ & \scriptsize ${\widetilde W}_{6}$  \\
\hline
10 & 0.99938 & 2.00113 & 2.99967 & 3.99968 & 5.00004 & 6.00009 \\
20 & 1.00008 & 2.00004 & 3.00002 & 4.00001 & 5.00002 & 5.99998 \\
40 & 1.00004 & 2.00007 & 3.00003 & 4.00013 & 5.00000 & 6.00012 \\
60 & 0.99999 & 2.00001 & 2.99998 & 4.00006 & 5.00001 & 6.00004 \\
100 & 1.00001 & 2.00001 & 3.00004 & 4.00002 & 5.00000 & 6.00002 \\
150 & 0.99997 & 2.00001 & 3.00002 & 4.00001 & 5.00004 & 6.00002 \\
\hline
\end{tabular}
}
\end{center}
\caption{Weight vectors for different bootstrap sample sizes without population noise}
\label{table:weight}
\end{table}

\begin{table}[h]
\begin{center}
\scalebox{0.90}{
\begin{tabular}{||c|c|c|c|c|c|c|c|c||}
\hline
Sample size & \scriptsize $\delta \widetilde{W}_{1}$ & \scriptsize $\delta \widetilde{W}_{2}$ & \scriptsize $\delta \widetilde{W}_{3}$ & \scriptsize $\delta \widetilde{W}_{4}$ & \scriptsize $\delta \widetilde{W}_{5}$ & \scriptsize $\delta \widetilde{W}_{6}$ \\
\hline
10 & 0.02039 & 0.03495 & 0.00725 & 0.01284 & 0.00934 & 0.00781  \\
20 & 0.00204 & 0.00118 & 0.00156 & 0.0016 & 0.00154 & 0.00217
     \\
40 & 0.00153 & 0.00316 & 0.003 & 0.00299 & 0.00155 & 0.00245
    \\
60 &  0.00077 & 0.00072 & 0.00076 & 0.00172 & 0.00056 & 0.00088
    \\
100 & 0.00087 & 0.00104 & 0.00106 & 0.00074 & 0.00092 & 0.00174  \\
150 & 0.00069 & 0.00052 & 0.00049 & 0.00034 & 0.0009 & 0.00079
     \\
\hline
\end{tabular}
}
\end{center}
\caption{Standard errors (SEs) of weight vectors for different bootstrap sample sizes without population noise}
\label{table:var-weight}
\end{table}

\begin{table}[h]
\begin{center}
\scalebox{0.90}{
\begin{tabular}{||c|c|c|c|c|c|c|c|c|c||}
\hline
Sample size & \scriptsize $t_{1}$ & \scriptsize $t_{2}$ & \scriptsize $t_{3}$ & \scriptsize $t_{4}$ & \scriptsize $t_{5}$ & \scriptsize $t_{6}$     \\
\hline
10 & 49.0036 & 57.25731 & 413.59599 & 311.55844 & 535.45297 & 768.20304  \\
20 & 490.03614 & 1697.92281 & 1924.45761 & 2506.22625 & 3237.56853 & 2759.18714 \\
40 & 653.08249 & 632.31813 & 1000.64225 & 1339.80026 & 3227.4301 & 2453.23403  \\
60 & 1291.6086 & 2787.94434 & 3970.08524 & 2321.44326 & 8930.72652 & 6798.62171 \\
100 & 1152.23006 & 1930.15583 & 2824.33706 & 5433.56226 & 5444.65661 & 3455.64034 \\
150 & 1457.42293 & 3855.09323 & 6125.15592 & 11916.92773 & 5580.49053 & 7587.86445 \\
\hline
\end{tabular}}
\end{center}
\caption{SEs of weight vectors for different bootstrap sample sizes without population noise}
\label{table:t}
\end{table}

A bootstrap sample is a sampled data set that is drawn with replacement from the original master population.
For our numerical demo, the master population data set has $N_{b} = 1024$ data records/rows, and we drew 1024 bootstrap samples, each of which has a much smaller chosen sample size. The regression weight vector learned by the proposed regression algorithm from the $1024$ bootstrap re-samples with the respective sample sizes $10$, $20$, $40$, $60$, $100$, and $150$ records. The zero bias term is guaranteed to be negligible from data standardization by subtraction from the sample mean.
Including additional columns from the response variable for the quantum encoding, we can emulate classically the quantum regression training with
$13$ qubits ~($2^{13} = 1024 \times 8$) without padding additional zeros.

Due to the variance of the bootstrap samples, the trained weight vectors fluctuate among these samples.
To establish our baseline errors from sampling and training, we show the ideal case with six features where the training data has no noise $\delta = 0$ to observe if we can emulate the learning.
As shown in Table~\ref{table:weight}, the training reproduces the theoretical values for the synthetic data we generate with the ideal weight vector $\overline{\bf W} = (1.0, 2.0, 3.0, 4.0, 5.0, 6.0)$ and we observe that the bootstrap sampling for the learning is reproduced for various batch sizes.
SEs of the weight vector $\delta {\bf \widetilde W} = (\delta \widetilde{W}_{1}, ~\delta \widetilde{W}_{2},....,\delta \widetilde{W}_{6})$ stay small and almost unchanged for distinct batch sizes as shown in Table~\ref{table:var-weight}. With the SEs in weight staying more or less constant, we expect a much larger $t$ for the features with higher weights.
If we look at the $t$-statistics metrics defined by the ratio
$t_{i} = {\widetilde W}_{i}/\delta {\widetilde W}_{i}$ as shown in Table~\ref{table:t}, we observe these values are much greater than one, representing the statistical significance of the learned results.
This shows that bootstrap sampling analysis is a valuable tool and generalizable in practice beyond the Gaussian noise hypothesis~\cite{Statistical Learning}, typically imposed in traditional statistical analysis.

To further confirm the practicality of the training approach with noise present in the map between features ${X_{i}}$ and the response variable $Y_{i}$, we go through the simulation with the noise level $\delta = 0.1$. For this case, we observe the
deviation of the learned weight vectors away from the ideal case without noise. With small sample sizes $10, 20$, and $40$, the sample mean weights can deviate from the theoretical weight vector more than what is indicated by the noise $\delta = 0.1$ in Table~\ref{table:weight-random}.
This is due to the sample variance being more pronounced at smaller batch sizes, as indicated in the noise-free case $\delta = 0$ in Table~\ref{table:var-weight}.

For larger sample sizes $60$, $100$, and $150$, we do observe that the mean weight vector from training mostly reproduces what is expected for the noise level
$\delta = 0.1$. The SEs of the weight vectors for the noisy cases are shown in Table \ref{table:var-weight-random}. When the learned weight vectors significantly deviate from theoretical values, we observe a corresponding larger SE for the weight vector. This correlation gives us guidance on how reliable our learned weight vectors are.
For example, for the first feature $\widetilde{W}_{1}$ with the sample size $20$, we see a large deviation from the theoretical value $1\pm 0.1$.
We also observe a larger deviation in its SE: $\delta \widetilde W_{1}$ at the sample size $20$.
In Table \ref{table:t-random}, we observe that the overall $t$ values are lower in comparison with the cases with no noise $\delta = 0$ due to the presence of non-sampling noise. In addition, we can identify that the overall $t$ values are the largest for the batch size $150$. This indicates that we
can use bootstrap sampling with the optimal sample size of $150$.

For larger batch sizes greater than $150$ (not shown), we start to observe the deviation from what we expect from theoretical values for the weight vector $\bf \overline W$.
This is because the ensemble training from the resampled data sets is under-fitting due to higher chances of duplicated data records in each sample, leading to training bias.
Even though the bias hinders us from drawing quantitative inferences from the data,
this behavior does not prevent us from selecting important features based on the $t$-statistics metrics defined by the ratio
$t_{i} = \widetilde{W}_{i}/\delta {\widetilde W}_{i}$ as shown in Table~\ref{table:t-random} and can be avoided with smaller bootstrap sample size. Note that this is also the case when there is no noise $\delta = 0$~(Table~\ref{table:t}), but occurs at a larger batch size $> 150$, not shown in Table~\ref{table:weight}.

\begin{table}[h]
\begin{center}
\scalebox{0.90}{
\begin{tabular}{||c|c|c|c|c|c|c|c|c|c||}
\hline
Sample size & \scriptsize ${\widetilde W}_{1}$ & \scriptsize ${\widetilde W}_{2}$ & \scriptsize ${\widetilde W}_{3}$ & \scriptsize ${\widetilde W}_{4}$ & \scriptsize ${\widetilde W}_{5}$ & \scriptsize ${\widetilde W}_{6}$  \\
\hline
10 &  1.24838 & 2.01457 & 2.92003 & 3.85253 & 4.69864 & 5.62674\\
20 & 0.02961 & 2.66497 & 1.64403 & 3.86877 & 5.15871 & 5.44711  \\
40 & 1.80006 & 0.74388 & 2.9371 & 3.97106 & 5.45333 & 5.65369\\
60 & 0.89376 & 1.83072 & 2.97385 & 4.07048 & 4.74884 & 5.72274 \\
100 & 0.63928 & 1.19575 & 3.04848 & 3.61821 & 5.09804 & 6.29805 \\
150 &  1.14361 & 1.79783 & 2.8523 & 3.95466 & 4.75087 & 5.82112 \\
\hline
\end{tabular}}
\end{center}
\caption{Weight vectors for different bootstrap sample sizes with Gaussian noise $\delta = 0.1$}
\label{table:weight-random}
\end{table}

\begin{table}[h]
\begin{center}
\scalebox{0.90}{
\begin{tabular}{||c|c|c|c|c|c|c|c|c||}
\hline
Sample size & \scriptsize $\delta \widetilde{W}_{1}$ & \scriptsize $\delta \widetilde{W}_{2}$ & \scriptsize $\delta \widetilde{W}_{3}$ & \scriptsize $\delta \widetilde{W}_{4}$ & \scriptsize $\delta \widetilde{W}_{5}$ & \scriptsize $\delta \widetilde{W}_{6}$   \\
\hline
10 &  10.3005 & 12.11829 & 13.45395 & 10.91457 & 7.85075 & 15.71212 \\
20 &  18.87171 & 28.18201 & 32.64924 & 14.18558 & 18.51711 & 21.57881 \\
40 & 16.54549 & 46.15104 & 6.43802 & 17.95377 & 20.97624 & 8.10855 \\
60 & 5.92399 & 5.39804 & 7.86823 & 6.12054 & 7.67524 & 5.49687 \\
100 & 9.85255 & 17.82527 & 8.84548 & 19.16585 & 10.84676 & 9.2562 \\
150 & 3.92479 & 3.14805 & 2.88643 & 3.30615 & 3.54041 & 3.47353 \\
\hline
\end{tabular}}
\end{center}
\caption{SEs of the weight vectors for different bootstrap sample sizes with Gaussian noise $\delta = 0.1$}
\label{table:var-weight-random}
\end{table}

\begin{table}[h]
\begin{center}
\scalebox{0.90}{
\begin{tabular}{||c|c|c|c|c|c|c|c|c|c||}
\hline
Sample size & \scriptsize $t_{1}$ & \scriptsize $t_{2}$ & \scriptsize $t_{3}$ & \scriptsize $t_{4}$ & \scriptsize $t_{5}$ & \scriptsize $t_{6}$     \\
\hline
10 &  0.1212 & 0.16624 & 0.21704 & 0.35297 & 0.5985 & 0.35811 \\
20 &  0.00157 & 0.09456 & 0.05035 & 0.27273 & 0.27859 & 0.25243 \\
40 & 0.10879 & 0.01612 & 0.45621 & 0.22118 & 0.25998 & 0.69725 \\
60 & 0.15087 & 0.33915 & 0.37796 & 0.66505 & 0.61872 & 1.04109 \\
100 &  0.06488 & 0.06708 & 0.34464 & 0.18878 & 0.47001 & 0.68041 \\
150 &  0.29138 & 0.57109 & 0.98818 & 1.19615 & 1.3419 & 1.67585\\
\hline
\end{tabular}}
\end{center}
\caption{Bootstrap-statistics metrics for the weight vectors for different bootstrap sample sizes with Gaussian noise $\delta = 0.1$}
\label{table:t-random}
\end{table}

\subsection{Feature importance and regularization}
In machine learning, we may have a potentially large list of features that can be used to describe the mapping between the response variable and input variables. Regularization provides an algorithmic way to select the optimal subset of original features quickly before performing a more detailed bootstrap sampling analysis for the finalized features. Regularization
penalizes the model with many features with important weights to avoid over-fitting the noise present in the training data, to induce more error in noisy hardware.
Since the regularization is done outside the quantum loop, this is a valid strategy for hybrid quantum machine learning. Here we demonstrate that the optimal feature selection can be enabled by turning on regularization in the cost function.
To establish the baseline, we generate synthetic data without Gaussian noise $\delta_{0} = 0$ where the response variable $Y=\sin( x)$ depends on the independent real variable $x$ in infinite order and the values are distributed randomly between the values $[-1,1]$. This is an infinite series for any real $x$ values, but can be truncated to a finite series when $x$ ranges between $[-1,1]$, which is the case for our normalized features. For regularization, we test with $L1$ regularization and $L2$ regularization.
We found the $L1$ regularization works robustly with the NM optimization algorithm in this case.

\begin{figure}[h]
    \centering
    \includegraphics[scale=0.75]{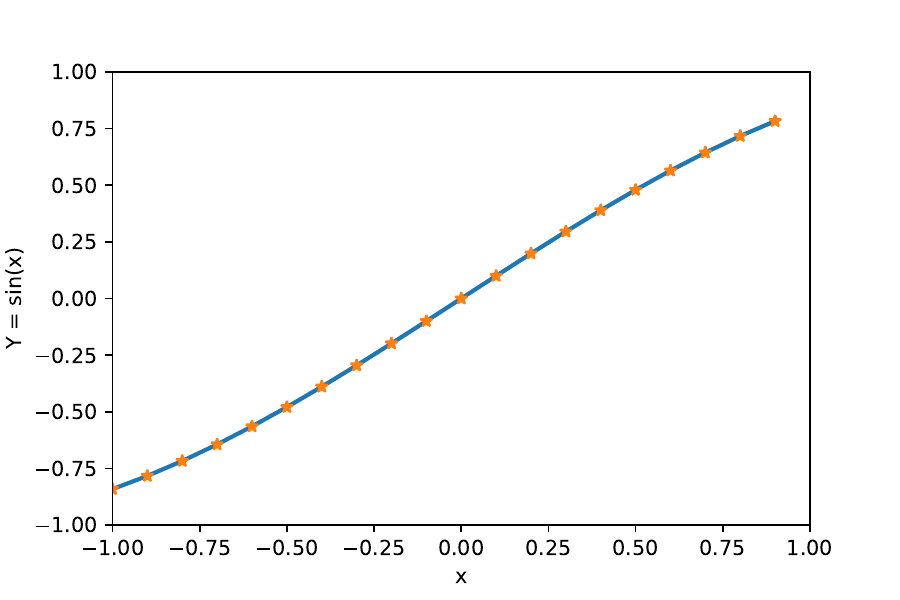}
    \caption{Prediction Agreement with the weights from theory results (solid line) and the weights from training from simulation (stars) are truncated to the fourth decimal accuracy: \\
    ${\bf {\widetilde W}}=(0.9992, 0.000001435, -0.1629, -0.000002908, 0.004134, -0.0004442, 0.000004760, \\ 0.0000012079, 0.000005189, 0.0003221, 0.0008932,0.0001738,0.000001149, -0.000001869, \\ -000001375)$.
    }
    \label{fig:agreement}
\end{figure}

In the following demonstration, we show that the nonlinear features can be built first in the feature space so that the linear regression algorithm can be used by feature preprocessing for the nonlinear regression model building.
We generate the synthetic data with controlled mapping between predictors~(features) ${\bf X} = (X_{1} = x, X_{2} = x^{2}, X_{3} = x^{3} ..., X_{15} = x^{15})$ and the target~(response) variable $Y \in {R} = \sin( x) = \sum_{n \in +\mathbb{Z}} (-1)^{n+1}{x}^{2n-1}/(2n-1)!$. The number of population records
$2^5$ are generated where each feature $X_{n}$ is uniformly generated between values $[-1,1]$. With $L1$ regularization, we use the very small
regularization parameter $\alpha = 1.2 \times 10^{-7}$, and alternating signs for the initial weight ansatz as $\sign(W_{1}, W_{3}, W_{5} .., W_{15}) = (+1, -1, +1,.., -1)$.
Our hybrid algorithm converges to the optimal weight parameter ${\bf \widetilde W} \approx {\bf \overline W}$, which selects the first few odd terms as the important features.
Noticed that the optimization may end up with a much smaller cost function, but with the wrong signs.
However, this confirms the experience that constraints from domain knowledge are typically required since the classical optimization algorithms can only be used as a filter for possible solutions, and even a mathematical global minimum solution may not be reasonable for the domain of applications.
What we found is that $L1$ regularization works with the proper regularization parameters $\alpha$. The number of vanishing weights in the converged weight vector reveals itself to select the feature effectively.
Taking the best-learned weight to produce the predicted value $Y$ for $x$ values ranging between $[-1, 1]$ as shown in Fig.~\ref{fig:agreement}, we can reproduce what we expected from synthetic training data.

\section{Conclusion}

We have presented the explainable quantum regression algorithm and detailed the plausible
implementation with quantum hardware with more connectivity. The algorithm is constructed exactly without
errors from approximate theory schemes such as Trotter errors or approximate circuit ansatz. In
addition, the model weights responsible for the explainability of the regression model are identified as
the cosine function of the variational rotation angles via controlled phase gates; the same controlled
phase gates can be used for the state preparation with larger gate complexity. We narrow the
gap between algorithm development and hardware implementation by this work to expedite the
realization in different hardware through the discussion in hardware implementation, typically
ignored in other algorithm work, by showing how the state vector evolution can be determined
by controlled phase gates with high connectivity precisely. The variational regression algorithm is most
likely to demonstrate optimal gate complexity in hardware with highly connected neutral cold atoms or cold ion systems, 
despite being realized first via a superconducting IQM machine~\cite{Frans}.

Although deterministic and non-uniform state preparation is still costly at large data limits
(see Appendix D), we mitigate this problem by replacing master data training with bootstrap samples with ensemble
regression model training from each sample, as demonstrated in the numerical results and recent experimental results (with gradient descent optimization) by the IQM
machine\cite{Frans}. This mitigates the notorious barren
plateau problem. In addition, we showed that a quantum regression map with an encoded data structure
can greatly reduce the time complexity compared to the classically constructed regression map.
Furthermore, we pointed out implementing nonlinear regression models with the linear regression
map but with preprocessed columns of nonlinear features, as shown in the last instance in the numerical results.
This reduces the circuit depth for the quantum nonlinear regression map in model training, useful for noisy hardware.

To conclude, we have constructed an explainable quantum regression algorithm and a new use
case for quantum machine learning without theory approximation errors. It would be interesting to
benchmark the algorithm with an optimal compiler in its respective quantum hardware to scrutinize
the potential quantum utilities.

\section{Acknowledgments}
We acknowledge Phil Lotshaw for his sincere feedback on the manuscript. C.-C. Joseph Wang and Ryan Bennink acknowledge the support by the DOE Office of Science, Office of ASCR, under FWP No. ERKJ354.

\appendix
\section{Proof of cost function from measurement}
\begin{equation}
\begin{aligned}
& |\Psi_{0}\rangle = \sum_{l',~m'} x_{l'm'}\cos{\phi_{m'}}|l'm'\rangle, \\
& \hat{M} = \sum_{l''}\sum_{m''}\sum_{m'''}|l''m''\rangle\langle l''m'''|, \\
& \hat{M}|\Psi_{0}\rangle = \sum_{l'}\sum_{l''}\sum_{m'}\sum_{m''}\sum_{m'''} x_{l'm'}\cos{\phi_{m'}}\langle l''m'''|l'm'\rangle |l''m''\rangle, \\
& = \sum_{l'}\sum_{m'}\sum_{m''}x_{l'm'}\cos{\phi_{m'}}|l'm''\rangle. \\
&~\rm Using~the~orthogonality~relation, \\
& \langle l''m'''|l'm'\rangle = \delta_{l''l'}\delta_{m'''m'} \\
& \langle \Psi_{0}| \hat{M} |\Psi_{0}\rangle =
\sum_{ll'}\sum_{m,~m',~m''} x_{lm}x_{l'm'}\cos{\phi_m}\cos{\phi_{m'}}\langle lm|l'm''\rangle \\
& = \sum_{l}\sum_{m}\sum_{m'} x_{lm}x_{lm'}\cos{\phi_{m}}\cos{\phi_{m'}} \\
& = \sum_{l}(\sum_{m} x_{lm}\cos{\phi_{m}})^{2}, \\
&~\rm Q.E.D.
\end{aligned}
\label{eq:cost function}
\end{equation}

\section{Success probability measurement}
By the measurement outcome from a perfectly trained model with less noisy data and good independent features selected in Eq.~(16),
we expect vanishing measurement results ${\rm Pr_{perfect}}=0$ from a perfect destructive interference, as the predicted response $\hat{y}_{l\in\{0,~1,~2,...,~L-1\}}$ agrees with the actual response $y_{l}$ compared to models that are not well trained.

For a poor model, what the model learned is the vanishing weights ${\overline W}_{m}$.
If the response variable is standardized with subtraction from its mean value, we expect the standardized bias term to be zero, and we expect a finite outcome from Eq.~(16) only contributed from the standardized response data $y^{S}_{l}$ as
\begin{equation}
{\rm Pr_{0}} \propto \sum \limits_{l}(y_l^{S}{\cos\phi_y})^{2}.
\end{equation}

For the worst model, we expect the signs of the weights to be all wrong and
a much larger probability outcome is expected as
\begin{equation}
{\rm Pr_{Worst}}\propto \sum\limits_{l}(2 y_l^{S}{\cos\phi_y})^{2}.
\end{equation}
We can define the goodness model metrics $G_{\rm M}$ for the trained model as ${G_{M}}\equiv 1-{\rm Pr_{M}}/{\rm Pr_{0}}$; equivalently, the coefficient of determination.
\begin{equation}
{G_{{\rm M}}}:\left\{
  \begin{tabular}{ll}
  =~$1$~ &~Perfect~ \\
  =~$0$~&~Poor~  \\
  = $-3$~&~Worst~~~~~~~~~~~~~~~~~~~~~~~.
  \end{tabular}
  \right.
\end{equation}
For the worst circumstance where the best estimate of the weight is wrong in signs,
the goodness of the worst model ${G_{\rm {Worst}}}$ will approach the value $-3$. This will be the case where the optimizer
is not set up correctly to find the minima, but finding the maxima or the sign for rotational angle for the response variable is not well taken off.
For meaningful training results, the goodness metrics should be in the range $G_{\rm M} \in (0,1]$.

Notice that the model metrics ${G_{{\rm M}}}$ are independent of any normalization convention in the algorithm.
To conclude whether the measurement result can be differentiated, we can estimate what is needed for the measured probability
$\rm{Pr_{0}}$ to be resolvable in experiments.
In terms of the standardized response variable $y_{l}^{S}$ after global normalization,
the probability $\rm Pr_{0}$
can be expressed as
\begin{equation}
  {\rm Pr_{0}} = \frac{1}{R^{2}}\sum\limits_{l}(y_l^{S}{\cos\phi_y})^{2},
\end{equation}
where $R^{2}$ is the global normalization factor after state preparation.
In terms of typical column standardized classical data,
the global norm $R^{2}$ is given by
\begin{equation}
  R^{2} = \sum\limits_{l}{y_{l}^{S}}^{~2} + \sum\limits_{lm}{x_{lm}^{S}}^{2}
  = \sigma_{y^{S}}^2 + \sum\limits_{m}{\sigma^{2}_{x_{m}^{S}}},
\end{equation}
in which the number of row is given by $L$, the sample variance for the input variable $x_{m}^{S}$ with column $m$ is given by $\sigma^{2}_{x_{m}^{S}}$ and the variance for the response variable $y$ is given by $\sigma^{2}_{y^{S}}$.
Finally, we can arrive at the following expression for the probability $\rm Pr_{0}$ as
\begin{equation}
\begin{aligned}
 & {\rm Pr_{0}} = \cos^{2}{\phi_y} \frac{\sigma_{y^{S}}^2}{\sigma_{y^{S}}^2+\sum\limits_{m}\sigma_{x_{m}^{S}}^2} \\
 & = \frac{1}{1 + F},
\end{aligned}
\end{equation}
in which $\phi_y=\pi$ is used and the total sample variance of all $M$ features is given by $\sum\limits_{m}\sigma_{x_{m}^{S}}^2$, and the relative sample variance ratio factor $F$ between all features and the response is defined by $F \equiv \sum\limits_{m}
{\sigma_{x_{m}^{S}}^{2}}/{\sigma^{2}_{y^{S}}}$.

For the relative variance factor $F$, it scales with the number of encoded features $M$ under the same standardization procedure before quantum data encoding.
Therefore, we expect the success probability for poor regression training ${\rm Pr_{0}}=\frac{1}{1+F}$ to scale inversely to the number of features $M$.
For a perfect model to be built, there should be a destructive interference that leads to zero observable probability ${\rm Pr}_{M}$ regardless
of the number of shots.
To be distinguishable from ${\rm Pr}_{0}$ in probability measurement error $\delta\epsilon$, it will take more than $(Pr_{0}(1-\delta\epsilon))^{-1} \propto (1+M)(1+\delta\epsilon)$ shots to observe the first expected success probability for small error $\delta\epsilon \ll 1$. 

Notice that the error $\delta \epsilon$ scales differently for different encoding schemes. For example,
assuming identical read-out error $\delta_{M}$ for a physical qubit states $'0','1'$, 
the read-out operator ${\hat R}_{0}$ for the '$0$' state is ${\hat R}_{0}= (1-\delta_{M})|0\rangle\langle 0|+ \delta_{M}|1\rangle\langle 1|$. 
For '$1$' state, the read-out error operator is $\hat R_{1}= \delta_{M}|0\rangle\langle 0| + (1-\delta_{M})|1\rangle\langle 1|$. Therefore, we can estimate the net encoded readout error in leading order in $\delta_{M}$
for one-hot encoding with $N_Q=L(M+1)$ physical qubits as
$\langle \Psi_{0}|{\hat M}{\hat R}_{0}^{\otimes N_{Q}-1} \otimes {\hat R}_{1}|\Psi_{0}\rangle-\langle \Psi_{0}|{\hat M}|\Psi_{0}\rangle=-N_{Q}\delta_{M}=-\delta \epsilon$, in which the minus sign represents the reduction in probability. For compressed encoding,
the error is exponentially suppressed to leading order in $\delta_{M}$ as given by
the expression $\langle \Psi_{0}|{\hat M}{\sum_{q,q'}\hat R}_{0}^{\otimes q} \otimes {\hat R}_{1}^{\otimes N_{L}-q}\otimes {\hat R}_{0}^{\otimes q'} \otimes {\hat R}_{1}^{\otimes N_{M}-q'}|\Psi_{0}\rangle-\langle \Psi_{0}|{\hat M}|\Psi_{0}\rangle=-(N_{L}+N_{Q})\delta_{M}=-\delta \epsilon $, in which $N_L = \lceil \log_2 L \rceil$ physical qubits and $N_M = \lceil \log_2 (M+1) \rceil$ physical qubits are used. We recapitulate the results in the abstract.

\section{Hardware implementation}
In cold-ion hardware, native gates include arbitrary one-qubit Pauli rotational gates and the two-qubit $\rm{XX}$ gate~\cite{Small Programmable Cold Ion}. The controlled phase (CPH) gate and the CNOT gates can be realized using the $XX$ gate in conjunction with Pauli 1-qubit rotations. The $H$ gate can be decomposed as $\rm{{R_{X}}(\pi){ R_{Y}}(\pi/2)}$ also on the platform. (Note that arbitrary 1-qubit gates plus any entangling 2-qubit gate constitute a universal set, so the Rydberg platform implements a universal set.)

In the Rydberg atom platform~\cite{Rydberg1}, any rotation in the Bloch sphere can be implemented, and the native two-qubit gate is $\rm{ZZ}$ type. The CNOT gate can also be decomposed in this platform as $\rm{(I \bigotimes H) C_{Z} ( I \bigotimes H)}$
where the controlled $\rm{Z}$ gate $\rm {C_{Z}}$, which is a special case for the controlled phase gate, is enabled by Rydberg states. The $\rm{CPH}$ gate can be decomposed in principle in terms of the $\rm{CNOT}$ gate with a one-qubit rotational gate~\cite{book}.
\subsection{One-hot encoding}
The specifics of the $\rm{CPH}$ gates vary with the encoding schemes.
For the one-hot amplitude encoding, the factorization of controlled two-body Pauli rotations $U_{C}^{j}$ along an axis is required.
Typically, this can be achieved in a preferred Pauli $\rm{Z}$ axis in a particular platform up to
a single qubit rotation from a native axis to the $\rm{Z}$ axis. For example, the native axis for cold ions would be Pauli $\rm{X}$
and the native axis for the Rydberg atom will be Pauli $\rm{Z}$.
The multi-qubit controlled phase gate can be implemented for the native Pauli $\rm{Z}$ axis as
$U_{C}^{m}=\rm{\prod_{j}\exp(-i\phi_{m}\frac{Z_{A}-I_{A}}{2}\otimes\frac{Z_{j}-I_{j}}{2})}$.
Equivalently, it can also be decomposed locally as
\begin{equation}
    U_{C}^{m}=\rm{\prod_{j}e^{-i\frac{\phi_{m}}{4} Z_{A}\otimes Z_{j}}
    e^{+i\frac{\phi_{m}}{4} Z_{A}\otimes I_{j}}
    e^{+i\frac{\phi_{m}}{4} I_{A}\otimes Z_{j}}
    e^{-i\frac{\phi_{m}}{4} I_{A}\otimes I_{j}}},
\end{equation}
in which the last unitary exponential factor is the idler unitary operator, which is state-independent and can be dropped. By digital 1-local and 2-local gate operation, the time complexity for each feature is $O(L)$. With $M$ features, the time complexity will be of $O(LM)$, the same as the time complexity with the state preparation as discussed in Sect. 3.~1.~1.
For partial globally-addressed analog gate operation $U_{C}^{m}$ on each feature,
the time complexity is greatly reduced to be $O(1)$ and scales as $O(M)$ in total. With a programmable, fully connected global analog gate with all features and the involved commutative Pauli operators,
$\prod_{m} U_{C}^{m}$ can be fused into a global unitary, and therefore, the time complexity can be minimized to
$O(1)$ at the expense of the time complexity $O(M)$ for the classical controls.
For the hardware with native $\rm{X}$ axis such as cold ions, we need to apply Pauli $\rm{Y}$ rotation $\rm{R_{Y_{A, j}}(-\pi/2)}$ to each physical qubit state in $U_{C}(\phi_{m})$ as
\begin{equation}
\begin{aligned}
&  U_{C}^{m} = \prod_{j} \rm{R_{Y_{A}}(+\pi/2)} \rm{R_{Y_{j}}(+\pi/2)} e^{-i\frac{\phi_{m}}{4} \rm{X_{A}\otimes X_{j}}}\rm{R_{Y_{j}}(-\pi/2)}\rm{R_{Y_{A}}(-\pi/2)}~\\
& \otimes \rm{R_{Y_{A}}(+\pi/2)}  e^{+i\frac{\phi_{m}}{4} \rm{X_{A}\otimes I_{j}}} \rm{R_{Y_{A}}(-\pi/2)} \\
& \otimes e^{+i\frac{\phi_{m}}{4} \rm{I_{A}\otimes X_{j}}},
 \end{aligned}
\end{equation}
in which the digital and local decomposition has arrived.
For the state preparation, the gate needs to be applied to each qubit reserved for the data registry.

\subsection{Compact binary encoding}
For the compact binary encoder, controlled phase gates are more complicated to implement. We suggest the application of a global entanglement gate, M{\o}lmer-S{\o}rensen ($\rm{MS}$) gate with an ancilla qubit, to achieve the quantum logic gate~\cite{Peter Zoller}.
The $\rm{MS}$ gate unitary operator in trapped cold ions is typically expressed as
\begin{equation}
    U_{MS}(\theta_{MS},\phi)=\rm{\exp(-i\frac{\theta_{MS}}{4}(\cos(\phi)S_{X}+\sin(\phi)S_{Y})^{2})},
\end{equation}
in which $\rm{S_{X,Y}=\sum_{i}{X_{i},\sum_{i} Y_{i}}}$ are the collective Pauli-operators.
With the help of an ancilla qubit, Pauli-string operation along a Pauli-axis $\rm{X, Y}$ can be enabled by choosing
the value and the sign of the phase $\phi$ to be ${0,\pi}$.
For $\phi =(0,\pi)$, $U_{MS} = (U_{MS}^{X}(\theta_{MS},\phi=0), U_{MS}^{Y}(\theta_{MS},\phi=\pi))$.
Angle $\theta_{MS}$ can be used to tune the rotation angle for the MS gate, where the exhaustive implementation is listed in Table $1$ in the reference. Notice that any discrepancy between the algorithms we develop can be easily adjusted to the native preferred axis for any platform after global rotation without many difficulties. The same comments hold as the one-hot encoder for the state preparation with controlled phase gates based on the MS gate. For Rydberg atoms, the research on $\rm{MS}$ gates for two atoms and multiple atoms is just in their infancy~\cite{Martin, Martin2, Lukin}.

For well-connected qubits, the multi-controlled phase gates would be sufficient to implement with only one ancilla qubit in principle, especially when the connectivity is close to infinitely long-ranged.  Partly, the practicality of the implementation is limited by the range of the phase gates that can be applied uniformly across physical qubits.
To add the finely programmable capability, individual and segmented digital addressing with the global entangled MS gate and local gates are
possible.
For example, for cold ions in a one-dimensional linear Paul trap and a two-dimensional Penning trap,
the ions are mostly uniformly distributed with long-range Ising interactions at the center of the trap. Therefore, it would be wise to select the ions away from the edges of the trap as the data register to reduce the sophistication of waveform engineering. Due to this controlled scalability limitation, we anticipate machine learning to be limited to a certain number of qubits, which limits the amount of training data that can be encoded for training concurrently.

\section{Resource Estimation}
Our time complexity analysis from the previous sections (Appendix C and Sect. 3) indicates that digital global gates reduce the gate time complexity compared to digital local gates.
For the one-hot encoder, with available local and global gates, the time complexity is $T_{O} \in O(LM+M)$ ($O(LM)$ from state preparation
and $O(M)$ from the quantum regression map; see Appendix C) with the memory footprint of $O(LM)$. The reduced time complexity comes from the encoded data structure.

For the compact binary encoder, the time complexity from  $T_{C} \in O(L^2M^2+M^2)$, including state preparation and regression map generation, respectively.
To decide the overall $Pr_{0}$ cost for the encoders at scale, we also need to consider factors beyond time complexity, such as the product of the qubit footprint (space complexity) and time complexity. The qubit $Pr_{0}$ required for the one-hot encoder $Q_{O}\in O(LM)$ is exponentially more costly than the $Pr_{0}$ for the compact binary encoder
$Q_{C} \in O(log_{2}LM)$.
Since we need to have a tall table ($L \gg M+1$), in which $M+1$ is the estimate for the Vapnik–Chervonenkis dimension for the linear regression model,  with a generalizable linear regression model without overfitting in classical learning theory (referring to LEARNING FROM DATA
by Y. S. Abu-Mostafa, M. Magdon-Ismail, and Hsuan-Tien Lin), 
the overall $Pr_{0}$ cost ratio $R_{C}/R_{O} = T_{C}Q_{C}/T_{O}Q_{O} \in O(log_{2}LM)$.
For classical algorithms, the widely accepted time complexity for the regression map is $O(LM^2)$ without consideration
of the matrix inversion time complexity, which is inferior to the quantum regression map with the embedded data structure.

However, the time complexity for classical algorithms for classical data preparation is $O(LM)$, the same as the one-hot encoder, but better than the compact binary encoder $O(L^2M^2)$, since there is no structural map to take
advantage of.
The total $Pr_{0}$ cost $R_{CL}$ for the classical computation that includes the memory cost $O(LM)$ for
the classical data would be $R_{CL} \in O(L^{2}M^{3})$, which shows a slight disadvantage over the resource cost $R_{O}\in O(L^2M^2)$
and $R_{C} \in O(L^{2}M^{2}log_{2}LM)$ without considering the matrix inversion, which is functionally equivalent to solving optimization in the hybrid algorithm.
\end{document}